\documentstyle[12pt]{article}   

\large
\newcommand{\tiN}{\raisebox{-6.5pt}{$\displaystyle
\stackrel{\displaystyle N}{\sim}$}}

\makeatletter
\@addtoreset{equation}{section}
\makeatother

\title{On the solution of the initial value constraints for general
relativity coupled to matter in terms of Ashtekar's variables}
\author{T. Thiemann\thanks{ithiemann@phys.psu.edu} \\
       Institute for Theoretical Physics, RWTH Aachen,\\
       D-52074 Aachen, Germany}
\date{{\small Preprint PITHA 93-1, January 93}}

\begin{document}

\maketitle                     

\begin{abstract}

The method of solution of the initial value constraints for pure canonical
gravity in terms of Ashtekar's new canonical variables due to CDJ
(see \cite{1}) is further developed in the present paper.
There are 2 new main results :

1) We extend the method of CDJ to arbitrary matter-coupling again for
non-degenerate metrics : the new feature is that the 'CDJ-matrix' adopts a
nontrivial antisymmetric part when solving the vector constraint and that the
Klein-Gordon-field is used, instead of the symmetric part of the CDJ-matrix,
in order to satisfy the scalar constraint.

2) The 2nd result is that one can solve the general initial value constraints
for arbitrary matter coupling by a method which is completely independent
of that of CDJ. It is shown how the Yang-Mills and gravitational Gauss
constraints can be solved explicitely for the corresponding electric fields.
The rest of the constraints can then be satisfied by using either scalar
or spinor field momenta.\\
This new trick might be of interest also for Yang-Mills theories on curved
backgrounds.

\end{abstract}

\section{Introduction}

The general solution of the initial value (iv) constraints with respect to
the vector and the scalar constraint within the Ashtekar-framework (see
\cite{3}) for pure gravity has a strikingly simple structure when restricting
it to non-degenerate metrics.\\
The gravitational action in the new canonical variables introduced
by Ashtekar (see ref. \cite{3}) has the form
\begin{equation}
^E S=\int_R dt\int_\Sigma d^3x \{P^a_i\dot{\omega}_a^i-
[-\omega_t^i (^E{\cal G}_i)+N^a(^E V_a)-\kappa\tiN(^E C)]\} \; .
\end{equation}
Here $P^a_i=-i/2\kappa\sqrt{\det(q)} e^a_i$ where $\kappa/(8\pi)$ is
Newton's coupling constant and $e^a_i$ is the triad of the metric
$q:=(q_{ab})$ on the hypersurface $\Sigma$ while $\omega_a^i$ is the
Ashtekar-connection. We use indices a,b,c,.. from the beginning
of the alphabet to denote valences of tensors defined on the initial data
hypersurface $\Sigma$ while indices i,j,k,.. from the middle of the alphabet
describe the O(3)-gauge group structure of a generalized tensor.
CDJ (see \cite{1}) have shown that the ansatz
\begin{equation}
P^a_i=\Psi_{ij}B^a_j  \; ,
\end{equation}
where $B^a_i:=\frac{1}{2}\epsilon^{abc}\Omega_{bc}^i$ is the magnetic field
with respect to the Ashtekar-connection, $\Omega_{ab}^i$ being its field
strength and $\epsilon^{abc}$ is the totally skew (metric-independent) tensor
density of weight 1, solves both the gravitational part of the vector
constraint
\begin{equation}
^E V_a:=\Omega_{ab}^i P^b_i\stackrel{!}{=}0
\end{equation}
and the gravitational part of scalar constraint
\begin{equation}
^E C:=\Omega_{ab}^i P^a_j P^b_k \epsilon_{ijk}\stackrel{!}{=}0
\end{equation}
provided the CDJ-matrix $\Psi$ is subject to the following conditions :
\begin{equation}
\Psi^T=\Psi\;\mbox{and}\;(tr(\Psi))^2-tr(\Psi^2)=0 \; .
\end{equation}
Here $A^T$ means the transpose of the matrix $A$. Inserting this ansatz into
the gravitational part of the Gauss-constraint one obtains when using the
Bianchi-identity (${\cal D}$ is the gauge covariant differential acting on
arbitrary generalized tensors)
\begin{equation}
^E{\cal G}_i:={\cal D}_a P^a_i=B^a_j{\cal D}_a\Psi_{ij}\stackrel{!}{=}0
\end{equation}
which is now a differential condition on $\Psi$ and cannot be solved by
purely algebraic methods any more. Up to now there are no solutions known to
equation (1.6) for full gravity.\\
Furthermore, it should be stressed that even if one had also the general
solution to equation (1.6) then one would only have obtained the constraint
surface of the 'non-degenerate sector' of the gravitational phase space, not
its reduced phase space since one did not factor by the gauge orbits yet.
Nevertheless, this method could be of some importance for obtaining the
reduced phase space for full gravity as might be indicated by the fact that
it {\em is} of some help in model systems (see \cite{4}).\\
In this paper we are going to discuss the iv constraints in terms of
Ashtekar's new variables for arbitrary matter coupling.
The analysis will be in the canonical framework as in equation (1.1), so all
the fields will be subject to 4 different types of constraints, namely \\
1) the gravitational Gauss constraint (Lagrange multiplier : $-\omega_t^i$)\\
2) the Yang-Mills Gauss constraint (Lagrange multiplier : $-A_t^i$)\\
3) the vector constraint (Lagrange multiplier : $N^a$, the shift vector)
and\\
4) the scalar constraint (Lagrange multiplier : $\tiN,\;\mbox{where}\;
N:=\sqrt{\det(q)}\tiN$ is the lapse function).\\
The matter sector which we are going to discuss consists of (we will not
dwell on how to derive the 3+1 form of the various actions, for details see
ref. \cite{5}; the 3+1 form of the Higgs-action is not derived there but
it can be obtained by a calculation similar to that for the Klein-Gordon
action so we can omit this here)
\begin{equation}
^{matter}S=^{KG}S+^W S+^C S+^{YM}S+^H S \; ,
\end{equation}
i.e. a collection of (real) Klein-Gordon-fields with arbitrary potential
$V(\phi)$ (we do not display the summation over the different scalar fields)
\begin{equation}
^{KG}S=\int_R dt\int_{\Sigma}d^3x \{\pi\dot{\phi}-[N^a\pi\phi_{,a}
+\frac{1}{2}\tiN(\pi^2+\det(q)(q^{ab}\phi_{,a}\phi_{,b}+V(\phi)))]\} \; ,
\end{equation}
a (collection of) (complex valued rather than Grassmann-valued) Weyl spinor
field(s) which couple only to the self-dual part of the spin-connection (this
is explained in refs. \cite{5} and \cite{6}; we write down only one spinor
field which may stand for an arbitrary number of Weyl-fields of possibly
both chiralities), including an arbitrary spinor potential (e.g. the usual
mass term when at least 2 spinor fields of both chiralities are present)
\begin{eqnarray}
^W S & = & \int_R dt\int_{\Sigma}d^3x \{\pi^T\dot{\psi}-[-\omega_t^i\pi^T
\tau_i\psi \nonumber\\
 & & +N^a\pi^T{\cal D}_a\psi-4\tiN(\kappa P^a_i\pi^T\tau_i{\cal D}_a\psi+
 \det(q) V(\psi^A,\pi_A))]\} \; ,
\end{eqnarray}
where $\tau_i=-i/2\sigma_i,\;\sigma_i$ the usual Pauli-matrices,
${\cal D}_a\psi:=[\partial_a+\omega_a^i\tau_i]\psi$ and $\pi=(\pi_A),\psi
=(\psi^A),
\; A=1,2$ SU(2) spinor indices, \\
a Yang-Mills-field for a semi-simple gauge group G (such that the
Cartan-Killing metric $(d_{IJ}), \; I,J,..$ internal indices of this gauge
group, is non-degenerate; if $T_I$ are the generators of the Lie algebra LG
of G then $d_{IJ}=tr(T_I T_J), \; [T_I,T_J]=f_{IJ}\;^K T_K$)
\begin{eqnarray}
^{YM} S & = & \int_R dt\int_\Sigma d^3x \{E^a_I\dot{A_a^I}-[-A_t^I{\cal D}_a
E^a_I\nonumber \\
& & +N^a F_{ab}^I E^b_I+\frac{g^2}{2}\tiN q_{ab} d^{IJ}(E^a_I E^b_J+B^a_I
B^b_J)]\} \; ,
\end{eqnarray}
where $E^a_I$ is the YM electric field, $B^a_I:=1/(2g^2)d_{IJ}\epsilon^{abc}
F^J_{bc},\; F^I_{ab}$
being the YM-field strength, is the YM magnetic field and g the YM coupling
constant, \\
a cosmological constant ($\Lambda$) term
\begin{equation}
^C S=\int_R dt\int_{\Sigma}d^3x \tiN \det(q) \Lambda \;
\end{equation}
and a Higgs-field for either the gravitational gauge group O(3) or the YM
gauge group G
\begin{eqnarray}
^{EH} S & = &\int_R dt\int_{\Sigma}d^3x \{\pi_i\dot{\phi}^i-[-\omega_t^i
\epsilon_{ij}\;^k\phi^j\pi^k+N^a\pi_i{\cal D}_a\phi^i \nonumber \\
 & + & \frac{1}{2}\tiN(\delta^{ij}\pi_i\pi_j+\det(q)(q^{ab}\delta_{ij}
 ({\cal D}_a \phi^i)({\cal D}_b\phi^j)+V(\phi^i)))]\} \; \mbox{or} \\
^{YMH} S & = &\int_R dt\int_{\Sigma}d^3x \{\pi_I\dot{\phi}^I-[-A_t^I
f_{IJ}\;^K\phi^J\pi^K+N^a\pi_I{\cal D}_a\phi^I \nonumber \\
 & + & \frac{1}{2}\tiN(d^{IJ}\pi_I\pi_J+\det(q)(q^{ab}d_{IJ}({\cal D}_a
 \phi^I)({\cal D}_b\phi^J)+V(\phi^I)))]\} \; .
\end{eqnarray}
After this brief introduction of our notation, we can outline the plan
of the paper : \\

In section 2 we will show how the method of CDJ can be extended to include
arbitrary physically relevant matter couplings. The deviations from the
source-free case are that the CDJ-matrix $\Psi$ fails to be symmetric and
that the scalar constraint is not solved for $\Psi$ but for one
Klein-Gordon-momentum $\pi$. The equation for $\pi$ is purely algebraic and
of 2nd or 4th order repectively depending on whether the YM field is coupled
or not. This is just sufficient in order that the scalar constraint be
solvable algebraically, for example, by the methods of Cardano and Ferrari
(see \cite{2}). As for the source-free case, this method does not solve the
gravitational Gauss constraint.

In section 3 we show that by employing a method which is completely
independent
of the CDJ-framework it is possible to solve {\em all} constraints of general
relativity coupled to arbitrary matter. An advantage of this method is that
the solutions of the constraint equations are {\ remarkably} simpler than
those obtained by the method presented in section 2.\\
First we show how the gravitational Higgs-field can be used to solve
both the gravitational Gauss constraint and the scalar constraint. The
equations for the Higgs field are at most quadratic. The YM-Higgs-field can
(in general) be used in order to satisfy the YM Gauss constraint. Now it is
the vector constraint which is not solvable by purely algebraic methods for
the gravitational field. However, by using again the gravitational Higgs
field and at least 3 Klein-Gordon fields one can solve both the vector and
the scalar constraint.\\
By a similar line of approach one can also solve only the YM Gauss-constraints
purely for the YM electric field. The rest of the constraints
can then be satisfied by purely algebraic methods if at least 4
Weyl spinor fields are present which, of course, corresponds to the physical
reality. This latter option has the advantage that one can forget about
scalar fields altogether which have not yet be proved to exist at all. \\
The results of section 3 may be of interest also for the canonical approach
to (pure) YM theory on curved or flat background metrics.\\

The paper concludes with an appendix in which some explicit formulas for
special cases of couplings, i.e. when the YM-field is absent, are given,
valid when applying the method of section 2. There we also display the
formulas of Cardano and Ferrari for the reader who wants to arrive at an
explicit solution of the initial value constraints including a YM-field when
applying the framework of section 2.

\section{Solving the vector and the scalar constraint}

The full vector constraint reads
\begin{equation}
V_a=\Omega_{ab}^i P^b_i+\pi\phi_{,a}+\pi^T{\cal D}_a\psi+F_{ab}^I E^b_I
+\pi_i{\cal D}_a\phi^i+\pi_I{\cal D}_a\phi^I \; .
\end{equation}
As CDJ (ref. \cite{1}) we make the ansatz
\begin{equation} P^a_i=\Psi_{ij} B^a_j \end{equation}
and obtain
\[ \Omega_{ab}^i P^b_i=\epsilon_{abc}B^c_i B^b_j \Psi_{ij}
=\det(B^a_i)\epsilon^{jik}B^k_a\Psi_{ij} \; ,\]
where $B^a_i B^i_b=\delta^a_b,\; B^a_i B^j_a=\delta^j_i$. This ansatz poses
no restrictions on $P^a_i$ as long as the magnetic fields are non-degenerate
(which however excludes flat space field configurations).\\
The important step is to decompose the CDJ-matix into its symmetric and
antisymmetric part
\begin{equation} \Psi:=S+A, \; S^T=S, \; A^T=-A \end{equation}
because only A enters the vector constraint which then can in fact be solved
for A : let $2\xi_k:=\epsilon_{ijk}A_{ij}, \; B:=\det(B^a_i)$, then
\begin{equation}
\xi_k=\pi[\frac{B^a_k}{2 B}\phi_{,a}]+[\frac{B^a_k}{2B}(\pi^T{\cal D}_a\psi
+F_{ab}^I E^b_I+\pi_i{\cal D}_a\phi^i+\pi_I{\cal D}_a\phi^I]
=:\pi\eta_k+\theta_k
\end{equation}
where the quantities $\eta_k,\theta_k$ do not depend on $\pi$.
Contracting equation (2.4) with $\epsilon_{ijk}$ we obtain
\begin{equation} A:=\pi T+R\; \mbox{where}\; T^T=-T,\; R^T=-R \; ,
\end{equation}
i.e. the matrix A is linear in $\pi$, homogenous only if the matter different
from the KG-field is absent.\\
We now insert this into the scalar constraint whose complete expression is
given by
\begin{eqnarray}
C & = & -\kappa\Omega_{ab}^i P^a_j P^b_k\epsilon_{ijk}+\frac{1}{2}[\pi^2
+\det(q)(q^{ab}\phi_{,a}\phi_{,b}+V(\phi))] \nonumber \\
  &  & -4(\kappa P^a_i \pi^T\tau_i{\cal D}_a\psi +\det(q) V(\psi^A,\pi_A))
+\Lambda\det(q)+\frac{g^2}{2}q_{ab}d^{IJ}(E^a_I E^b_J+B^a_I B^b_J)\nonumber\\
  &  & +\frac{1}{2}[\delta^{ij}\pi_i\pi_j+\det(q)
(\delta_{ij}q^{ab}({\cal D}_a\phi^i)({\cal D}_b\phi^j)+V(\phi^i))]\nonumber\\
  & & +\frac{1}{2}[d^{IJ}\pi_I\pi_J+\det(q)
  (d_{IJ}q^{ab}({\cal D}_a\phi^I)({\cal D}_b\phi^J)+V(\phi^I))] \nonumber \\
  & =: & ^E C+^{KG} C+^W C+^C C+^{YM} C+^{EH} C+^{YMH} C \; .
\end{eqnarray}
When inserting for $P^a_i$ the various terms involve different powers of A
which we now discuss in sequence :
\begin{equation}
-\frac{1}{\kappa}^E C=\epsilon_{abc}B^c_i B^a_m B^b_n
\epsilon^{ijk}\Psi_{jm}\Psi_{kn}
=B\epsilon_{mni}\epsilon^{jki}\Psi_{jm}\Psi_{kn}=B((tr(\Psi))^2-tr(\Psi^2))
\end{equation}
Since $tr(A)=tr(A^T)=-tr(A)=0,\; tr(S A)=tr(A^T S^T)=-tr(A S)=-tr(S A)=0$ for
any symmetric matrix S and any antisymmetric matrix A, it follows easily that
\begin{equation} ^E C=-\kappa B((tr(S))^2-tr(S^2)-tr(A^2)) \; ,
\end{equation}
such that $\pi$ enters $^E C$ only purely quadratically without a linear
term.\\
Let $\phi_{,a}B^a_i:=v_i$, then
\begin{equation}
\det(q)q^{ab}\phi_{,a}\phi_{,b}=-4\kappa^2 P^a_i P^b_i \phi_{,a}\phi_{,b} \; ,
\end{equation}
but $P^a_i\phi_{,a}=(S_{ij}+R_{ij})v_j$ because $T_{ij}v_j
=\epsilon_{ijk}v_j v_k/(2 B)\phi_{,a}=0$.
Hence
\begin{equation}
\det(q)q^{ab}\phi_{,a}\phi_{,b}=-4\kappa^2 v^T((S+R)^T(S+R))v
=-4\kappa^2 tr((S^2-R^2+[S,R])v\otimes v)
\end{equation}
which is independent of $\pi$.\\
Now
\begin{eqnarray}
& & \det(\det(q)q^{ab})=(\det(q))^3(\det(q))^{-1}=(\det(q))^2\nonumber \\
& = & -(2\kappa)^2\det(P^a_i P^b_i)=-(2\kappa\det(P^a_i))^2
\end{eqnarray}
from which follows (up to a sign) that
\begin{equation} \det(q)=2i\kappa B\det(\Psi) \; . \end{equation}
Since a term cubic in A enters $det(\Psi)$ only through $det(A)$ (this
follows
from the fact that $\Psi=A+S$ and that the determinant is a totally skew
multilinear functional) which vanishes in 3 dimensions, $det(q)$ is also only
quadratic in $\pi$. The explicit formula
\footnote{This formula can also be obtained by applying the theorem of
Hamilton-Cayley as was pointed out to the author by Ted Jacobson in private
communication} is given by
\begin{eqnarray}
\det(\Psi) & = & \frac{1}{3!}\epsilon^{ijk}\epsilon_{lmn}
(A+S)_{il}(A+S)_{jm}(A+S)_{kn} \nonumber \\
 & = & \frac{1}{3!}\epsilon^{ijk}\epsilon_{lmn}(S_{il}S_{jm}S_{kn}
 +3S_{il}S_{jm}A_{kn}+3S_{il}A_{jm}A_{kn}+A_{il}A_{jm}A_{kn}) \nonumber \\
 & = & \det(S)+\det(A)+\frac{1}{2}3!\delta_{[i}^l\delta_j^m\delta_{k]}^n
 (S_{il}S_{jm}A_{kn}+S_{il}A_{jm}A_{kn}) \nonumber \\
 & = & \det(S)+tr(SA^2)-\frac{1}{2}tr(S)tr(A^2) \; .
\end{eqnarray}
Altogether we have therefore
\begin{eqnarray}
^{KG} C & = & \frac{1}{2}[\pi^2-4\kappa^2 tr((S^2-R^2+[S,R])v\otimes v)
\nonumber \\
 & & +2i\kappa B(\det(S)+tr(SA^2)-\frac{1}{2}tr(S)tr(A^2))V(\phi))] \; .
\end{eqnarray}
The Higgs-sector differs from the KG-sector as far as the appearence of $\pi$
is concerned only in that $T_{ij}B^a_j\phi_{,a}=0, \;\mbox{while}\;
T_{ij}B^a_j{\cal D}_a\phi_k,T_{ij}B^a_j{\cal D}_a\phi_k\not=0$.\\
Let $v^i_j:=B^a_j{\cal D}_a\phi^i, \;v^I_j:=B^a_j{\cal D}_a\phi^I$ then
\begin{eqnarray}
^{EH} C & = & \frac{1}{2}[\delta^{ij}\pi_i\pi_j-4\kappa^2\delta_{ij}
tr((S^2-A^2+[S,A])v^i\otimes v^j) \nonumber \\
& &+2i\kappa B(\det(S)+tr(SA^2)-\frac{1}{2}tr(S)tr(A^2))V(\phi^i))]
\nonumber\\
^{YMH} C & = & \frac{1}{2}[d^{IJ}\pi_I\pi_J-4\kappa^2d_{IJ}
tr((S^2-A^2+[S,A])v^I\otimes v^J) \nonumber \\
& & +2i\kappa B(\det(S)+tr(SA^2)-\frac{1}{2}tr(S)tr(A^2))V(\phi^I))] \; .
\end{eqnarray}
There is only a quadratic appearence of $\pi$ again.\\
The contribution by the YM-sector however gives rise to a quartic constraint
for $\pi$ :\\
We have
\begin{eqnarray}
& & q_{ab} = \det(q)E^i_a E^i_a=\frac{(2\kappa)^4}{4\det(q)}\epsilon_{acd}
\epsilon_{bef}\epsilon^{ijk}\epsilon^{imn} P^c_j P^d_k P^e_m P^f_n\nonumber\\
& = &\frac{(2\kappa)^4 B^2}{2\det(q)}\epsilon_{rst}\epsilon_{uvw}
\delta_{[j}^m
\delta_{k]}^n \Psi_{jr}\Psi_{ks}\Psi_{mu}\Psi_{nv}B_a^t B_b^w \nonumber \\
 & = & \frac{(2\kappa)^4  B^2}{2\det(q)}3!\delta_{[r}^u\delta_s^v
 \delta_{t]}^w (\Psi^T\Psi)_{ru}(\Psi^T\Psi)_{sv} B_a^t B_b^w \nonumber \\
 & = & \frac{(2\kappa)^4 B^2}{2\det(q)}[(tr(\Psi^T\Psi))^2-tr((\Psi^T\Psi)^2)
 +2(\Psi^T\Psi)^2-2tr(\Psi^T\Psi)\Psi^T\Psi]_{tw} B_a^t B_b^w
\end{eqnarray}
Since $\Psi^T\Psi=(S-A)(S+A)=S^2-A^2+[S,A], \; tr([S,A])=0$, we conclude that
\begin{eqnarray}
q_{ab} & = & \frac{(2\kappa)^4 B^2}{2\det(q)}[(tr(S^2-A^2))^2-tr((S^2-A^2
+[S,A])^2)+2(S^2-A^2+[S,A])^2 \nonumber \\
       & & -2tr(S^2-A^2)(S^2-A^2+[S,A])]_{tw} B_a^t B_b^w \; .
\end{eqnarray}
As long as $\det(q)\not =0$ one is allowed to multiply the scalar constraint
by $\det(q)$ so that it becomes 4th order in $\pi$ because the contributions
of the other fields are at most quadratic in $\pi$ and $\det(\Psi)$ is,
according to formula (2.13), also only quadratic in $\pi$.\\
The final step is to collect all terms and to determine the coefficients
of the various powers of $\pi$. We have
\begin{eqnarray}
\det(\Psi) & = & [\det(S)+tr(SR^2)-\frac{1}{2}tr(S)tr(R^2)]
+\pi[tr(S(RT+TR)) \nonumber \\
& & -\frac{1}{2}tr(S)tr(TR+RT)]+\pi^2[tr(ST^2)
-\frac{1}{2}tr(S)tr(T^2)] \nonumber \\
           & =: & (a\pi^2+b\pi+c)/(2i\kappa B)
\end{eqnarray}
and
\begin{eqnarray}
\Psi^T\Psi & = & \{S^2-R^2-[S,R]\}-\pi\{RT+TR-[S,T]\}
-\pi^2 T^2 \; , \nonumber \\
(\Psi^T\Psi)^2 & = & [(S^2-R^2-[S,R])^2] \nonumber \\
& & -\pi[(RT+TR-[S,T])(S^2-R^2-[S,R])+(S^2-R^2-[S,R]) \nonumber \\
& & (RT+TR-[S,T])]-\pi^2[(RT+TR-[S,T])T^2 \nonumber \\
& & +T^2(RT+TR-[S,T])-(S^2-R^2-[S,R])^2] \nonumber \\
&  & +\pi^3[(RT+TR-[S,T])T^2+T^2(RT+TR-[S,T])]+\pi^4 T^4  \nonumber \\
& =: &  A_4\pi^4+A_3\pi^3+A_2\pi^2+A_1\pi+A_0
\end{eqnarray}
such that
\begin{eqnarray}
& & q_{ab}\frac{2\det(q)}{(2\kappa)^4 B^2}B_i^a B_j^b \nonumber \\
& = & [(tr(S^2-R^2))^2-4\pi tr(S^2-R^2)tr(RT)-2\pi^2(tr(S^2-R^2)tr(T^2)
\nonumber \\
& & -2(tr(RT))^2)+4\pi^3tr(RT)tr(T^2)+\pi^4(tr(T^2))^2-(tr(A_4)\pi^4
 +tr(A_3)\pi^3 \nonumber \\
 & & +tr(A_2)\pi^2+tr(A_1)\pi+tr(A_0))+2(A_4\pi^4+A_3\pi^3+A_2\pi^2+A_1\pi
 +A_0) \nonumber \\
& &  -2(tr(S^2-R^2)(S^2-R^2-[S,R])-\pi(tr(S^2-R^2)(RT+TR-[S,T]) \nonumber \\
& & +2tr(RT) (S^2-R^2-[S,R])) \nonumber \\
& & -\pi^2(tr(S^2-R^2)T^2+tr(T^2)(S^2-R^2-[S,R])-2tr(RT)(S^2-R^2
-[S,R])) \nonumber \\
 & & +\pi^3(2tr(RT)T^2+tr(T^2)(S^2-R^2-[S,R]))+\pi^4tr(T^2)T^2)]_{ij} \; .
\end{eqnarray}
We first collect the powers of $\pi$ contained in the non-YM part of the
scalar constraint and then multiply by $\det(q)$.
\begin{eqnarray}
& & ^E C+^{KG} C+^W C+^C C+^{EH} C+^{YMH} C \nonumber \\
& = & \{-\kappa B((tr(S))^2-tr(S^2)-tr(R^2)-2\kappa^2 tr((S^2-R^2+[S,R])v
\otimes v) \nonumber \\
&  & +c\frac{1}{2}V(\phi)-4(\kappa B^a_j(S+R)_{ij} \pi^T\tau_i{\cal D}_a\psi
+ c V(\psi^A,\pi_A))+c\Lambda  \nonumber \\
& &     + \frac{1}{2}[\delta^{ij}\pi_i\pi_j-4\kappa^2\delta_{ij}
tr([S^2-R^2+[S,R]]v^i\otimes v^j)   \nonumber \\
& &     +c V(\phi^i))]+\frac{1}{2}[d^{IJ}\pi_I\pi_J-4\kappa^2d_{IJ}
tr([S^2-R^2+[S,R]]v^I\otimes v^J)+c V(\phi^I))] \}   \nonumber \\
& + & \{2\kappa B tr(RT)+\frac{b}{2}V(\phi)-4(\kappa B^a_j T_{ij} \pi^T
\tau_i{\cal D}_a\psi+b V(\psi^A,\pi_A)) \nonumber \\
& &+b\Lambda+\frac{1}{2}[4\kappa^2\delta_{ij} tr([RT+TR-[S,T]]]v^i\otimes
v^j)+b V(\phi^i))] \nonumber \\
& &     + \frac{1}{2}[4\kappa^2d_{IJ} tr([RT+TR-[S,T]]v^I\otimes v^J)
+b V(\phi^I))]\}\pi      \nonumber \\
& + & \{\kappa B tr(T^2)+\frac{1}{2}[1+a V(\phi)]-4a V(\psi^A,\psi_A)
+a\Lambda    \nonumber \\
& & +\frac{1}{2}[4\kappa^2\delta_{ij} tr(T^2 v^i\otimes v^j) \nonumber \\
& & +a V(\phi^i)]+\frac{1}{2}[4\kappa^2d_{IJ} tr(T^2 v^I\otimes v^J)
+a V(\phi^I)] \}\pi^2 \nonumber \\
& =: & d+e\pi+f\pi^2
\end{eqnarray}
Hence
\begin{equation}
\det(q)[C-^{YM} C]=af\pi^4+(ae+fb)\pi^3+(ad+fc+eb)\pi^2+(bd+ec)\pi+cd \; .
\end{equation}
Finally we have thus, using $M^{ij}:=g^2 B^2(2\kappa)^4/4 B^i_a B^j_b d_{IJ}
(E^a_I E^b_J+B^a_I B^b_J)$
as an abbreviation
\begin{eqnarray}
& & \det(q)C \nonumber \\
& = &\{(tr(S^2-R^2))^2 tr(M)-tr(A_0)tr(M)+2tr(A_0 M)-2tr(S^2-R^2)\nonumber\\
& &   tr((S^2-R^2-[S,R])M)+cd\} \nonumber \\
& + & \{-4tr(S^2-R^2)tr(RT)tr(M)-tr(A_1)tr(M)+2tr(A_1 M)
+2(tr(S^2-R^2) \nonumber \\
& & tr((RT+TR-[S,T])M)+2tr(RT)tr((S^2-R^2-[S,R])M))+bd+ec\}\pi \nonumber \\
& + & \{-2(tr(S^2-R^2)tr(T^2)-2(tr(RT))^2)tr(M)-tr(A_2)tr(M)
+2tr(A_2 M) \nonumber \\
& &   -2(tr(S^2-R^2)tr(T^2 M)+tr(T^2)tr((S^2-R^2-[S,R])M) \nonumber \\
& &   -2tr(RT)tr((S^2-R^2-[S,R])M))+ad+fc+eb\}\pi^2 \nonumber \\
& + & \{4tr(RT)tr(T^2)tr(M)-tr(A_3)tr(M)+2tr(A_3 M) \nonumber \\
& &-2(2tr(RT)tr(T^2 M)+tr(T^2)tr((S^2-R^2-[S,R])M))+ae+fb\}\pi^3\nonumber\\
& + & \{(tr(T^2))^2tr(M)-tr(A_4)tr(M)+2tr(A_4 M)-2tr(T^2)
tr(T^2 M)\}\pi^4 \nonumber \\
& =: & \{(tr(T^2))^2tr(M)-tr(A_4)tr(M)+2tr(A_4 M)-2tr(T^2)
tr(T^2 M)\} \nonumber \\
 & &  [\pi^4+\alpha\pi^3+\beta\pi^2+\gamma\pi+\delta] \; .
\end{eqnarray}
The coefficients $\alpha,\beta,\gamma,\delta$ can now directly be plugged
into the resolution formulas for a general quartic equation due to Cardano
and Ferrari as outlined in the appendix.\\
We have thus arrived at the general solution of the initial value constraints
for arbitrary matter coupling and nondegenerate 3-metrics and magnetic fields
as far as the vector and scalar constraint are concerned. The matrix A and
the field $\pi$ are expressed in terms of the other fields. Note that one
cannot solve the scalar constraint for one of the 6 independent components
of the matrix S which appears in {\em fifth} order due to the term cd in
(2.22) (according to Galois theory (see \cite{2}) the general
algebraic equation of order larger than 4 is not solvable by radicals).\\
In order to complete
the solution of the iv constraints one would like to solve the
Gauss-constraint
in terms of the matrix S which however becomes now highly nontrivial and is
beyond the scope of the present paper.\\
In the appendix we give the above solution for special cases i.e. when the
Yang-Mills field is absent. In particular, if one only couples the
Klein-Gordon
field, then the scalar constraint depends only on $\pi^2$ and the above
formulas simplify tremendously.

\section{Solving all constraints}

\subsection{Solution by using the scalar fields}

First of all we show how a Higgs-field can be used for any semi-simple
gauge group with rank $\ell$ to satisfy the Yang-Mills gauge constraint
(note that for semisimple groups $f_{IJK}:=f_{IJ}\;^L d_{LK}$ is totally
skew; see ref.
\cite{7} for the necessary Lie algebra terminology).\\
The YM Gauss constraint is
\begin{equation}
{\cal G}_I:={\cal D}_a E^a_I+f_{IJ}\;^K\phi^J\pi_K\stackrel{!}{=}0 \; .
\end{equation}
Let V be the complex vector space of dimension dim(G) which is the
representation
space of the adjoint representation of LG. Let further $Y:=\phi^I T_I,
\;(T_I)_{JK}=-f_{IJK}$
and $V^\perp(\phi^I)=\{X\in LG; \; ad(Y) X:=[Y,X]=0\}$ which shows that
$V^\perp$
depends only on $\phi$ (even the dimension k of $V^\perp$ depends,
in general, on
the specific element Y : choose a Cartan subalgebra $\cal H$ of LG and
choose the corresponding Weyl canonical form. If $Y\in\; \cal H$
then $1\le dim(V^\perp)=\ell$, if Y is a nonzero-root vector relative to
$\cal H$ then $dim(V^\perp)$ is not characterizable by $\ell$ only but will
depend on the algebra and the root chosen. One only knows then that
$dim(V^\parallel)\ge \ell+1$, see ref. \cite{7}, where $V^\parallel
:=V-V^\perp$).\\
We will regard the fields as Lie algebra valued by the identification
$\phi^I T_I=:\phi$ etc. \\
Contracting equation (3.1) with $X^I\in V^\perp$ yields
\begin{equation} X^I{\cal D}_a E^a_I=0  \end{equation}
because $X^I f_{IJ}^K\phi^J\pi_K=tr(T_K[X,\phi])\pi^K=tr(\pi[X,\phi])=0$ for
any $\pi$.
Hence the internal-vector density ${\cal D}_a E^a_I$ is weakly 'orthogonal'
to the subspace $V^\perp$. \\
Hence, the part of the Gauss-constraint which is 'parallel' to $V^\perp$
has to be satisfied independently of the momentum $\pi_I$. We will satisfy
the
constraint eqns. (3.2) which can be read as a condition on $E^a_I$ by
employing {\em the following new trick which is at the heart of the present
approach} : \\
Let
\begin{equation} E^a_I:=\epsilon^{abc}{\cal D}_b v_{cI} \end{equation}
where the generalized tensor $v_{aI}$ is yet arbitrary. Then,
using the torsion-freeness of the (purely metric-determined)
Riemann-connection $\Gamma^a\;_{bc}$ which acts on tensor indices only
(the torsion due to the spinor fields
shows up in $\omega_a^i$, more precisely in the spin-connection,
and is part of the reality conditions, see refs. \cite{5},\cite{6}) we have
\begin{equation} X^I{\cal D}_a E^a_I=g^2 f_{IJ}\;^K X^I B^{aJ}v_{a_K}=0 \;
\forall X\in V^\perp \end{equation}
which is a {\em purely algebraic} restriction for $v_{aI}$ and which can be
satisfied identically by solving for k components of $v_{aI}$
in terms of the others and of the magnetic as well as the Higgs fields. We
assume that this has been done in the following. The electric fields remain,
however, independent of the Higgs-momenta which is important for the sequel.
\\
Are there any restricions implied on the Lie-algebra valued 2-form
$t_{abI}:=E^c_I\epsilon_{cab}$ by representing it as a exterior differental
of the Lie-algebra valued 1-form $v_{aI}$ i.e. $t_I={\cal D}\wedge v_{I}$ ?
The purely algebraic properties of both generalized tensors are the same,
however one has to worry about the 'generalized integrability conditions'
obtained by taking the exterior differential of the last equation
\[ {\cal D}\wedge t_I=\frac{1}{2}f_I\;^{JK}F_J\wedge v_K \; . \]
The latter equation can now only be satisfied for an arbitrary $t_I$ if the
magnetic fields $F_I$ are non-degenerate. This restriction does not directly
show up in equation (3.3) because even for $A_I=0$ the rhs of eq. (3.3) is
nonvanishing for a suitable choice of $v_I$. Hence we obtain the same
restriction as
CDJ on the magnetic fields in order that the new trick works, at least
when there is a non-trivial Yang-Mills-potential. \\
We decompose the dual internal vector density $\pi_I$ into a part 'parallel'
and 'orthogonal' to $V^\perp$
\begin{equation} \pi_I:=\pi^\perp_I+\pi^\parallel_I \; , \end{equation}
where $\pi^\perp\in\; V^\perp,\; \pi^\parallel\in\; V^\parallel$
are dual internal vector densities. Hence it follows when inserting into the
Gauss-constraint
\begin{equation} {\cal D}_a E^a_I+Y_I^J\pi^\parallel_J=0 \; , \end{equation}
where it is understood that $E^a_I$ is replaced by equation (3.3). Equation
(3.6) can be solved for $\pi^\parallel$ since Y is regular on $V^\parallel$
while $\pi^\perp$ remains unspecified.\\
Hence we can regard the gravitational and YM-Gauss law as identically
satisfied in terms of 2 of the gravitational and dim(G)-k components of the
YM Higgs field and 1 component of the gravitational and k components of the
YM
electric field. Note, however, that there is the additional spin-density
$\pi^T\tau_i\psi$ contained in the gravitational Gauss constraint which upon
solving the Gauss constraint becomes part of the vector $v_{ai}$.\\
When inserting these solutions
into the vector constraint, one obtains again a genuinely differential
relation
between the relevant momenta $v_{ai}, \; v_{aI}$ and thus the vector
constraint fails to be algebraically solvable in terms of the gravitational
field. However, we can make use of the
gravitational field $\pi^\perp_i$ and/or the YM field $\pi^\perp_I$ and 3
Klein-Gordon momenta $\pi_\alpha\, \alpha=1,2,3$ in order to solve the rest
of the constraints by algebraic methods (the usage of scalar fields in order
to solve all constraints of GR plus scalar matter is most convenient in the
old ADM variables).
One solves first the 3 vector constraints in terms of $\pi_\alpha$ which will
then depend linearly on the fields $\pi^\perp_i, \;\pi^\perp_I$. Hence, the
scalar constraint also depends
only quadratically on $\pi^\perp_i, \;\pi^\perp_I$. Accordingly, by coupling
suitable matter, {\em it turns out to be possible to solve all the constraints
explicitely}. So matter helps to solve the complete iv constraints with
this method.\\
We will now give the explicit formulas. Let $\pi^\perp_i=:h_E \phi_i$
and $|\phi|^2:=\phi^i\phi^i$. The vector constraint reads
\begin{eqnarray}
V_a & = & [\Omega_{ab}^i P^b_i+F_{ab}^I E^b_I+\pi_I{\cal D}_a\phi^I+\frac{1}
{|\phi|^2}\epsilon_{ijk}\phi_j({\cal D}_b P^b_k)({\cal D}_a\phi^i) \nonumber \\
 & & +\pi^T{\cal D}_a\psi]+h_E\phi^i{\cal D}_a\phi^i+\pi_\alpha
 \phi^\alpha_{,a}
\end{eqnarray}
where it is understood that $\phi_I,E^a_I,P^a_i$ are expressed in terms of
the
other fields as derived above, the crucial point being that they do not
depend
on $h_E,\pi_\alpha$. We write this as a matrix relation
\begin{equation} \vec{u}_0+h_E\vec{u}+P(\pi_1,\pi_2,\pi_3)^T=0 \end{equation}
where the matrix P, which consists of the covariant derivatives of the fields
$\phi^\alpha$, is in general non-singular so that equation (3.8) can be
inverted to give
\begin{equation} \pi_\alpha=-(P^{-1})_\alpha^\beta[\vec{u}_0+h\vec{u}]_\beta
=:(a+b h_E)_\alpha \; . \end{equation}
Here the coefficients $a_\alpha,b_\alpha$ do not depend on $h_E$.\\
We finally insert this into the scalar constraint (again it is to be
understood that one has to insert the above solutions for $P^a_i,
E^a_I,\phi^I$ and that $q_{ab}$ is expressed in terms of $P^a_i$) and obtain
\begin{eqnarray}
C  & = & \{-\kappa\Omega_{ab}^i P^a_j P^b_k\epsilon_{ijk}+\frac{1}{2}
[a_\alpha^2+\det(q)(q^{ab}\phi^\alpha_{,a}\phi^\alpha_{,b}+V(\phi))]
+\Lambda\det(q) \nonumber \\
  & & -4(\kappa P^a_i \psi^T\tau_i{\cal D}_a\psi+\det(q) V(\psi^A,\pi_A))
  +\frac{g^2}{2}q_{ab}d^{IJ}(E^a_I E^b_J+B^a_I B^b_J) \nonumber \\
& & +\frac{1}{2}[(\frac{1}{|\phi|^2}\epsilon_{ijk}\phi_j({\cal D}_b P^b_k))^2
+\det(q)(\delta_{ij}q^{ab}({\cal D}_a\phi^i)({\cal D}_b\phi^j)+V(\phi^i))]
\nonumber \\
  & & +\frac{1}{2}[d^{IJ}\pi_I\pi_J+\det(q)(d_{IJ}q^{ab}({\cal D}_a\phi^I)
  ({\cal D}_b\phi^J)+V(\phi^I))]\} \nonumber \\
  & + & \{a_\alpha b_\alpha\}h_E+\{\frac{1}{2}b_\alpha^2+\frac{1}{2}|\phi|^2
  \}(h_E)^2 \nonumber \\
  & =: & \alpha (h_E)^2+\beta h_E+\gamma
\end{eqnarray}
which yields a quadratic equation for $h_E$. Hence, the inclusion of a
gravitational and YM Higgs-field together with this new method enables one
to obtain the general solution of the initial value constraints for arbitrary
matter coupling. Note that the new method introduced is completely
independent of the CDJ-framework since the CDJ-matrix does not enter the game
at any stage.\\

\subsection{Solution without using the scalar fields}

The method of section 3.1 (as well as of section 2) has the unattractive
feature that explicit use was made of the (possibly spurious) scalar fields.
We now show that one can do without them altogether by essentially the same
trick if one uses Weyl fields to solve the vector and scalar constraint as
well as the gravitational Gauss constraint. \\
Inserting the basic ansatz (3.3) into the Gauss constraint yields the purely
algebraic relation
\begin{equation} {\cal D}_a E^a_I=g^2 f_{IJ}\;^K B^{aJ}v_{a_K}
=-f_{IJ}\;^K\phi^J\pi_K \; \forall X\in V^\perp \end{equation}
for $v_{aI}$. There are $3\times dim(G)$ independent components of the
electric field contained in $v_{aI}$ and dim(G) constraints, so we can
choose dim(G) of the components of $v_{aI}$ to depend on the others as well
as on the magnetic fields, the Higgs field and the $V^\parallel$ part of the
Higgs-momenta. We assume that this has been done from now on.\\
Now one inserts formula (3.11) into the rest of the constraints. We assume
that at least 4 spinor fields are present and solve these remaining 7
constraints in terms of 7 components of spinor field momenta by using similar
algebraic methods as in eqn. (3.8) ff. If the spinor momenta enter the scalar
constraint only linearly,
i.e. $V(\psi^A,\pi_A)=V(\psi^A)$, there is not even the need to solve a
quadratic equation which simplifies the relevant formulas tremendously. Hence,
one could then write the vector and the scalar constraint as {\em one} matrix
relation as in eqn. (3.1.8) for the 7 components of spinor momenta chosen and
solve for them by methods of linear algebra. If this is not the case, then we
require that the spinor
potential is of at most 4th order in the spinor momenta (without spatial
derivatives) in order that the scalar constraint be solvable by algebraic
methods.\\
We refrain from giving the explicit formulas because the procedure how to
get them is identically the same as the one of section (3.1).\\
\\
Note : in case of the gravitational Gauss constraint one can interprete the
trick, eqn. (3.3), in the following nice way :\\
Provided that the magnetic fields are non-degenerate we can write
$v_{ai}=m_{ab}B^b_i, \; m_{ab}$ being an arbitrary tensor density of weight
-1. Then the Gauss-law yields
\begin{equation}
{\cal D}_a P^a_i=\epsilon_{ijk} B^a_j B^b_k m_{ab}=B B_c^i\epsilon^{abc}
m_{ab} =-\epsilon_{ijk}\phi^i\pi_k
\end{equation}
such that the antisymmetric part of the tensor density m is expressible in
terms of the gravitational Higgs-field
\begin{equation}
\epsilon^{abc}m_{ab}=-\frac{1}{B}B^c_i\epsilon_{ijk}\phi^i\pi_k
\end{equation}
while its symmetric part remains unspecified. Unfortunately, the symmetric
part enters the vector and scalar constraint differentiated which excludes
the possibility to solve for $m_{(ab)}$, at least by algebraic methods.\\
\\
\\

{\large Acknoledgements}\\
\\
I thank Professor Ted Jacobson for interesting discussions at the University of
Maryland and on line and for stimulating this work. He also pointed out the
restrictions on the magnetic fields for the ansatz (3.3).

\begin{appendix}

\section{Appendix}

\subsection{The resolution formula for quartic equations}

For the benefit of the reader we include here a brief recipe for solving
the general quartic equation. The formulas were first found by Cardano and
Ferrari (see \cite{2}).\\
1) Turn the general form of a quartic equation
\begin{equation} x^4+\alpha x^3+\beta x^2+\gamma x+\delta=0 \end{equation}
into its normal form by substituting $x=y-\alpha/4$ :
\begin{equation} y^4+p y^2+q y+r=0 \end{equation}
2) Make the ansatz
\begin{equation} y^4+p y^2+q y+r=(y^2+P)^2-(Qy+R)^2 \end{equation}
This factorizes into a product of two quadratic equations and can be solved
by standard methods. Comparison of coefficients yields
\begin{equation} p=2P-Q^2,\; q=-2QR,\; r=P^2-R^2 \end{equation}
and results in the so-called cubic resolvent
\begin{equation} P^3-\frac{1}{2}p P^2-rP+\frac{1}{8}q^2+\frac{1}{2}rp
=:P^3+a P^2+b P+c=0 \;. \end{equation}
3) Solve the cubic equation which upon substitution $P=t-a/3$ adopts its
normal form $t^3+et+f=0$. Let $D:=(e/3)^3+(f/2)^2$ (the discriminant) and
\begin{equation} u:=\sqrt[3]{-\frac{f}{2}+\sqrt{D}},\; v:=\sqrt[3]
{-\frac{f}{2}-\sqrt{D}} \end{equation}
then the 3 roots are given by
\begin{equation}
t_1=u+v, \; t_2=-\frac{u+v}{2}+i\sqrt{3}\frac{u-v}{2}, \;
t_3=-\frac{u+v}{2}-i\sqrt{3}\frac{u-v}{2}
\end{equation}
4) Use one of the roots $t_1,t_2,t_3$ of the cubic resolvent to determine
$P,Q,R$ and proceed with formula (A.3).\\
Since already $\alpha,\beta,\gamma,\delta$ look horrible when expressed in
terms of the independent components of the fields (compare formula (2.23))
we refrain from giving x in terms of $\alpha,\beta,\gamma,\delta$
explicitely
and rather discuss a feasible example.

\subsection{The extended method of CDJ applied to a special case}

First we switch off only the Yang-Mills field (and, necessarily, its
associated Higgs-field). Then the scalar constraint reduces to (recall
formula (2.21))
\begin{equation} f\pi^2+e\pi+d=0 \end{equation}
where $E^a_I=A_a^I=\pi^I=\phi^I=0$, also in the expressions for a,b,c -
compare formula (2.18). This is now only a quadratic equation for $\pi$.\\
Although the degree of the scalar constraint cannot be lowered further
without switching off the Klein-Gordon field altogether, a tremendous
simplification
occurs when one retains only the gravitational and the KG field because then
the matrix R vanishes. We obtain
\begin{equation}
\det(\Psi)=[\det(S)]+\pi^2[tr(ST^2)-\frac{1}{2}tr(S)tr(T^2)]=:(a\pi^2+c)
/(2i\kappa B)
\end{equation}
i.e. b=0. The scalar constraint reduces to
\begin{eqnarray}
C & = &  -\kappa B((tr(S))^2-tr(S^2)-\pi^2tr(T^2)) \nonumber \\
& & +\frac{1}{2}[\pi^2-4\kappa^2 tr(S^2 v\otimes v)+(a\pi^2+c)V(\phi))]
\nonumber \\
  & = &  \{-\kappa B(((tr(S))^2-tr(S^2))+\frac{1}{2}
  [-4\kappa^2 tr(S^2 v\otimes v)+c V(\phi)]\} \nonumber \\
  & &    + \{\kappa B tr(T^2)+\frac{1}{2}[1+a V(\phi)]\}\pi^2 \nonumber \\
  & = &  f \pi^2+d
\end{eqnarray}
i.e. e=0, the momentum $\pi$ enters the scalar constraint without a linear
term ! The matrix T is simply $T_{ij}=B^a_k/(2 B)\phi_{,a}\epsilon_{ijk}$.

\end{appendix}


\begin{thebibliography}{999}

\bibitem{1}  R.\ Capovilla, T.\ Jacobson and J.\ Dell,
             Phys. Rev. Lett. {\bf 63}(1989)2325; \\
             R.\ Capovilla et al.,   Class. Quantum
             Grav. {\bf 8}(1991)41;\\
             R.\ Capovilla et al.,
             Class. Quantum Grav. {\bf 8}(1991)57
\bibitem{2}  I.N.\ Bronstein, K.A.\ Semendjajew, Taschenbuch der Mathematik,
             23 rd edition (Harri Deutsch, Thun and Frankfurt/Main, 1987)
\bibitem{3}  A.\ Ashtekar,   Phys.\ Rev.\ {\bf D36} (1987)1587; \\
             A.\ Ashtekar,  New Perspepectives in Canonical Gravity
             (Monographs and Textbooks in Physical Science, Bibliopolis,
             Napoli, 1988);\\
             A.\ Ashtekar, Lectures on Non-Perturbative Canonical Gravity
             (World Scientific, Singapore, 1991)
\bibitem{4}  T.\ Thiemann, H.A.\ Kastrup, Nucl. Phys.  (1993)
\bibitem{5}  A.\ Ashtekar et al.,
             Phys.Rev. {\bf D40} (1989)2572
\bibitem{6}  T.\ Jacobson, Class. Quantum Grav. {\bf 5} (1988)L143
\bibitem{7}  S.\ Helgason, Differential Geometry, Lie Groups and Symmetric
             Spaces (Academic Press, San Diego, 1978)



\end{thebibliography}
\end{document}